\documentclass{article}
\usepackage{subcaption}  

\usepackage{ijcai26}

\usepackage{times}
\usepackage{soul}
\usepackage{url}
\usepackage[hidelinks]{hyperref}
\usepackage[utf8]{inputenc}
\usepackage{graphicx}
\usepackage{amsmath}
\usepackage{amsthm}
\usepackage{amsfonts,amssymb}
\usepackage{booktabs}
\usepackage{algorithm}
\usepackage{algorithmic}
\usepackage[switch]{lineno}
\usepackage{multirow}
\usepackage[table]{xcolor}
\usepackage{colortbl}

\title{SLSREC: Self-Supervised Contrastive Learning for Adaptive Fusion of Long- and Short-Term User Interests}

\author{
Wei Zhou$^1$
\and
Yue Shen$^1$
\and
Junkai Ji$^1$
\and
Yinglan Feng$^1$
\and
Xing Tang$^2$
\and
Xiuqiang He$^2$
\and
Liang Feng$^3$
\and
Zexuan Zhu$^1$
\\
\affiliations
$^1$School of Artificial Intelligence, Shenzhen University, Shenzhen 518060, China\\
$^2$Shenzhen Technology University, Shenzhen, China\\
$^3$College of Computer Science, Chongqing University, Chongqing, China\\
\emails
jerryzhou@szu.edu.cn, 
shenyue2023@email.szu.edu.cn, 
jijunkai@szu.edu.cn, 
yinglfeng@szu.edu.cn, 
liangf@cqu.edu.cn, 
xing.tang@hotmail.com, 
he.xiuqiang@gmail.com, 
zhuzx@szu.edu.cn
}
\begin{document}

\maketitle

\begin{abstract}

User interests typically encompass both long-term preferences and short-term intentions, reflecting the dynamic nature of user behaviors across different timeframes. The uneven temporal distribution of user interactions highlights the evolving patterns of interests, making it challenging to accurately capture shifts in interests using comprehensive historical behaviors. To address this, we propose \textbf{SLSRec}, a novel \textbf{S}ession-based model with the fusion of \textbf{L}ong- and \textbf{S}hort-term \textbf{Rec}ommendations that effectively captures the temporal dynamics of user interests by segmenting historical behaviors over time.
Unlike conventional models that combine long- and short-term user interests into a single representation, compromising recommendation accuracy, SLSRec utilizes a self-supervised learning framework to disentangle these two types of interests. A contrastive learning strategy is introduced to ensure accurate calibration of long- and short-term interest representations. Additionally, an attention-based fusion network is designed to adaptively aggregate interest representations, optimizing their integration to enhance recommendation performance. Extensive experiments on three public benchmark datasets demonstrate that SLSRec consistently outperforms state-of-the-art models while exhibiting superior robustness across various scenarios.We will release all source code upon acceptance.
\end{abstract}

\section{Introduction}

Recommender systems infer user preferences from historical interaction data in order to deliver personalized content. \cite{chen2022intent,tanjim2020attentive,zhou2023dynamic}. However, user interests in real-world scenarios often evolve at different temporal scales, typically consisting of relatively stable long-term preferences and dynamic short-term intentions. For example, on e-commerce platforms, users may exhibit persistent interests in digital products such as smartphones and laptops, while temporarily showing short-term interests in items like suitcases or chargers due to specific travel needs. Therefore, effectively modeling and disentangling users’ long- and short-term (LS-term) interests is critical for improving recommendation accuracy.

To address the temporal dynamics of user behaviors, sequential recommendation (SR) methods 
\cite{hidasi2015session,tang2018personalized,zhou2019deep,cen2020controllable,chang2021sequential}
have been proposed to model users’ evolving preferences from historical interaction sequences. 
By capturing temporal dependencies among user behaviors, SR models have achieved remarkable progress in personalized recommendation, especially in scenarios where user interests change over time.
Despite these advances, most existing SR models still face fundamental challenges in effectively modeling LS-term interests. Many approaches either focus on local sequential patterns to emphasize recent behaviors, or encode the entire interaction history into a single unified representation. 
Such designs make it difficult to distinguish stable long-term preferences from transient short-term intentions, thereby limiting their ability to capture multi-scale interest dynamics underlying complex user behaviors.

The above studies reveal a key insight: user behaviors is jointly driven by long-term preferences and short-term interests. Accordingly, effectively fusing LS-term interest modeling has become a central research focus in sequential recommendation. To this end, several methods \cite{an2019neural,lv2019sdm,yu2019adaptive,zhao2018plastic} attempt to integrate users’ long-term preferences and short-term intentions within a unified framework. Typically, these approaches employ collaborative filtering techniques (e.g., matrix factorization \cite{an2019neural,zhao2018plastic}) to model long-term interests, while utilizing sequential models such as LSTM \cite{lv2019sdm,yu2019adaptive,zhao2018plastic} and GRU \cite{an2019neural} to capture short-term dynamics.

However, despite modeling both LS-term interests, the absence of explicit supervisory signals makes it difficult to effectively calibrate their representations, limiting the models’ ability to accurately reflect users’ true preferences. To address this issue, \cite{zheng2022disentangling} introduces contrastive learning to supervise LS-term embeddings. Nevertheless, due to its reliance on overall historical behavior modeling, this approach still struggles to accurately capture users’ immediate needs, particularly when short-term interests change rapidly.

Broadly speaking, disentangling LS-term interests in sequential recommendation faces several key challenges. \textbf{(1) Temporal segmentation for multi-scale interest modeling:} User behaviors are unevenly distributed over time, with temporally proximate interactions exhibiting stronger correlations. Modeling user behavior over coarse historical sequences often fails to capture such temporal variations, motivating the need for session-level segmentation to better characterize interest evolution. \textbf{(2) Disentangled representation of long- and short-term interests:} Long-term interests reflect users’ stable preferences, while short-term interests correspond to immediate needs. Effectively modeling these two aspects requires separate yet complementary representations that capture both global stability and local sensitivity. \textbf{(3) Self-supervised calibration of LS-term interests:} In the absence of explicit labels, designing effective supervisory signals to distinguish and calibrate LS-term interest representations remains a challenging problem.

To address the above challenges, we propose \textbf{SLSRec}, a contrastive learning-based sequential recommendation framework that explicitly disentangles LS-term interests through session segmentation. Specifically, SLSRec consists of four key components: a long-term interest encoder, a short-term interest encoder, a contrastive learning module, and an adaptive fusion network. We first partition users’ historical interaction sequences into multiple sessions based on timestamps to capture temporal dependencies more effectively. The long-term interest encoder models stable user preferences by capturing session-level interest evolution, while the short-term interest encoder focuses on extracting dynamic intention shifts from recent interactions. Furthermore, we introduce a self-supervised contrastive learning strategy to explicitly calibrate LS-term interest representations. Finally, an attention-based fusion network adaptively aggregates LS-term interests for accurate interaction prediction.
In summary, the main contributions of this work are as follows:
\begin{itemize}
    \item We propose a session-based partition strategy to facilitate the disentanglement of LS-term interests, where long-term interests are modeled via session-level interest evolution and short-term interests are extracted from recent interactions.
    \item We design a contrastive learning framework to explicitly supervise LS-term interest representations and introduce an attention-based fusion network to adaptively aggregate LS-term interests.
    \item We conduct extensive experiments on three publicly available datasets, and the results demonstrate that SLSRec consistently outperforms state-of-the-art sequential recommendation methods.
\end{itemize}


\section{Related Work}

\subsection{LS-Term Interest Modeling}

In recent years, several approaches have attempted to jointly model users’ LS-term interests by combining collaborative filtering (CF) techniques with sequential recommendation models \cite{an2019neural,lv2019sdm,yu2019adaptive,zhao2018plastic,ying2018sequential,zheng2022disentangling}. These methods typically rely on CF-based models (e.g., matrix factorization) to capture long-term preferences, while employing sequential architectures to model short-term dynamics. Representative examples include hierarchical attention-based models \cite{ying2018sequential}, adaptive fusion mechanisms \cite{yu2019adaptive}, and more recent extensions incorporating contrastive learning or memory-based structures \cite{guo2024information,wei2023multi,zheng2022disentangling}. A detailed review of these methods is provided in the appendix.

Despite their effectiveness, most existing approaches lack explicit disentanglement between LS-term interest representations. The interactions between different temporal interests are often implicitly modeled, which may result in representation entanglement and insufficient semantic separation \cite{locatello2019challenging}. Moreover, their reliance on overall historical behavior modeling limits their ability to accurately capture users’ immediate needs, particularly when short-term interests evolve rapidly.
\subsection{Contrastive Learning for Recommendation}
Self-supervised learning, especially contrastive learning, has recently been adopted in sequential recommendation to enhance representation learning by exploiting intrinsic correlations in user behavior sequences.
Representative methods such as CLS4Rec \cite{xie2022contrastive} and COTREC \cite{xia2021self} extract self-supervised signals through sequence-level augmentations, improving robustness without explicit labels.
Subsequent studies further combine contrastive objectives with graph structures or cross-view perturbations to improve generalization \cite{wei2021contrastive,cai2023lightgcl,zhou2023equivariant}.Recent efforts also enrich contrastive supervision by incorporating semantic information or alternative sequence views.
UDA4SR \cite{shih2025recommendation} generates alternative behavior sequences to alleviate data sparsity, while SRA-CL \cite{cui2025semantic} constructs contrastive pairs via semantic retrieval in embedding spaces.

Beyond robustness improvement, contrastive learning has been shown to facilitate the decoupling of users’ LS-term interests by constructing informative positive and negative pairs, making it a promising tool for calibrating LS-term interest representations in sequential recommendation.

\section{Methodology}

\begin{figure*}[ht]
    \centering
    \includegraphics[width=0.87 \linewidth]{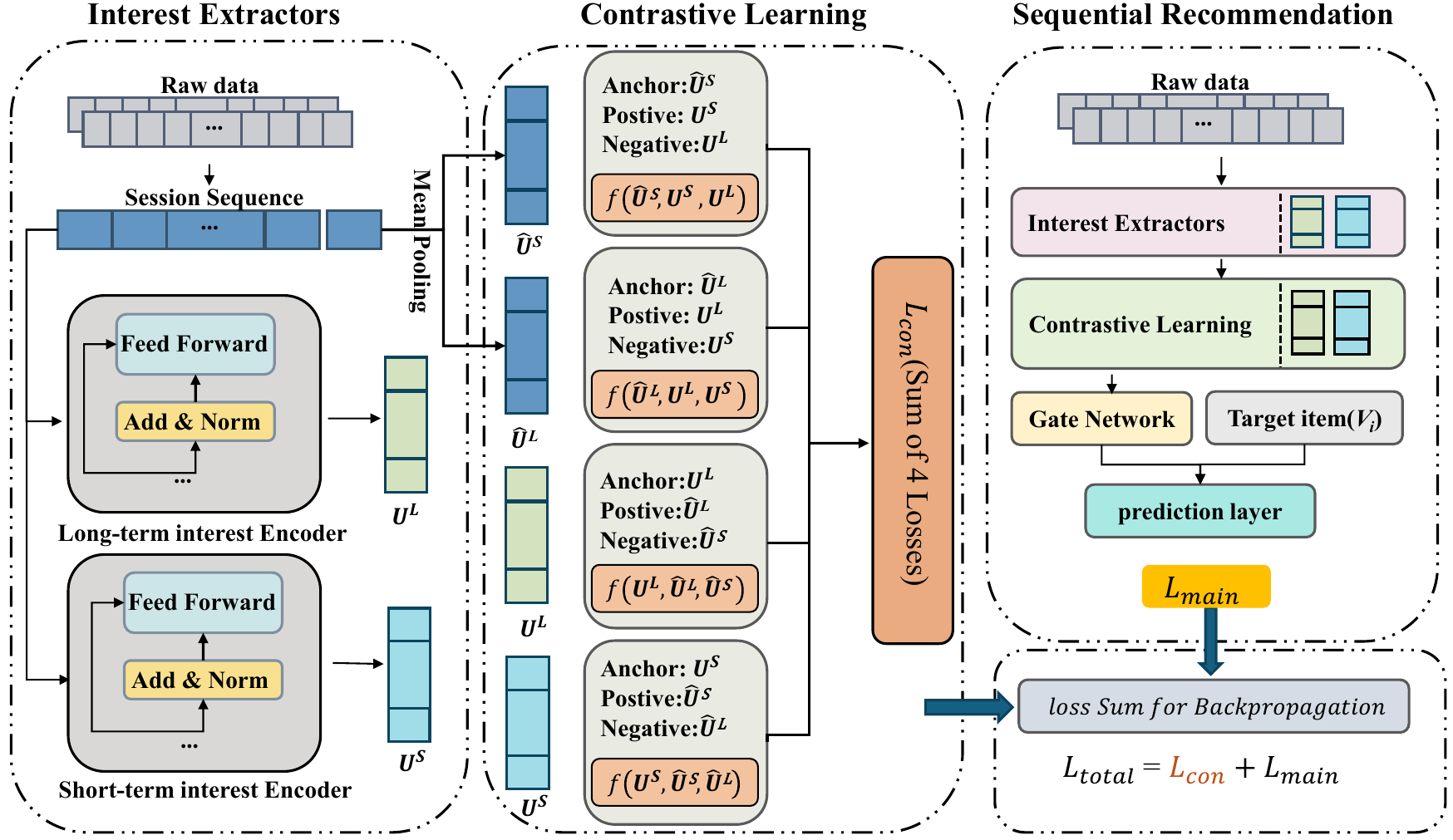}
    \centering
    \caption{The overall architecture of SLSRec}\label{fig:fig1}
\end{figure*}

In this section, we elaborate on the proposed contrastive learning-based framework for LS-term interests fusion. The overall architecture of SLSRec is illustrated in Fig.~\ref{fig:fig1}. In this paper, we use the term \emph{user behaviors} to denote timestamp-ordered user--item interactions. First, the user behavior sequence is segmented into multiple sessions by a time-aware session segmentation layer, taking into account the multi-session structure of historical interactions. 
Second, LS-term interest encoders are employed to model user interests at different temporal scales, which are explicitly supervised through a contrastive learning mechanism. 
Finally, a fusion prediction layer aggregates user interests across temporal scales and estimates the interaction probability with the target item $\mathbf{v}_{T}$. 
The details of each component are described as follows.

\subsection{Time-aware Session Segmentation Layer}
To model temporal dynamics in user behaviors, we partition historical interaction sequences into multiple sessions. User behaviors exhibit uneven temporal distributions: interactions occurring close in time are more correlated, while those separated by long intervals are less related. Session segmentation therefore enables high intra-session coherence while preserving long-term interest evolution across sessions, facilitating fine-grained temporal modeling.

We first construct an item embedding matrix $\mathbf{E} \in \mathbb{R}^{V \times d}$, where $V$ is the number of items and $d$ is the embedding dimension. Each interaction at time step $t$ is embedded as $\mathbf{v}_t \in \mathbb{R}^{d}$, forming a behavior sequence $\mathbf{S} \in \mathbb{R}^{s \times d}$. A time-aware segmentation function $\operatorname{TSD}(\cdot)$ divides $\mathbf{S}$ into sessions using a threshold $\omega$: two consecutive interactions are assigned to the same session if their time interval $\Delta t < \omega$, and to different sessions otherwise.
As a result, the behavior sequence $\mathbf{S}$ is partitioned into $k$ sessions:
\begin{equation}
[\mathbf{S}^{(1)}, \mathbf{S}^{(2)}, \dots, \mathbf{S}^{(k)}]
= \operatorname{TSD}(\mathbf{S}, \omega),
\end{equation}
where $\mathbf{S}^{(n)} \in \mathbb{R}^{l \times d}$ denotes the embedding matrix of the $n$-th session, $l$ is the maximum session length, and $k$ is the total number of sessions.

\subsection{Short-term Interest Encoder}
We apply the same attention encoder used for long-term interest modeling to capture the short-term interest in the current session $\mathbf{S}^{(k)}$, which is represented as the user's current interest representation.
\begin{equation}
\mathbf{u}_s = \operatorname{Attention}\!\left(\mathbf{S}^{(k)}\right).
\end{equation}

\begin{figure}[h]
	\center
	\includegraphics*[width=0.44\textwidth]{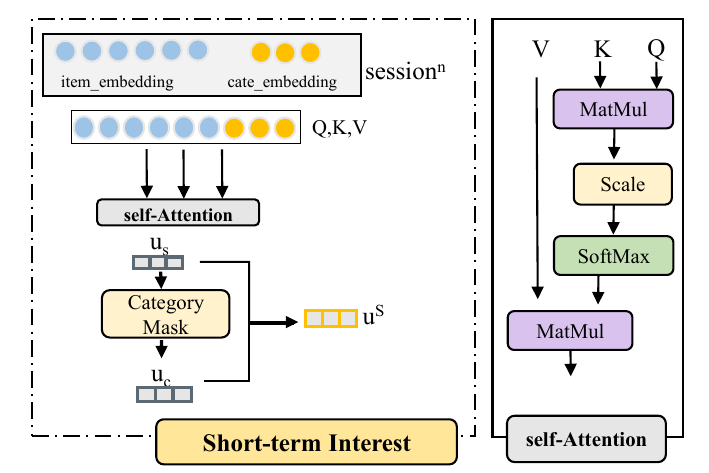}
	\centering
	\caption{Short-term interest encoder}\label{fig:fig3}
\end{figure}
We utilize a category mask matrix $\mathbf{M}$ in order to more accurately capture the user's short-term interest category features, we perform fine-grained modeling of short-term interest representations by combining category information, which in turn highlights the user's key interests in the current session. Specifically, given the high consistency of items within the same session, we extract the category sequence of each item from the current session and prioritize the category with the highest number of interactions. In the case of ties, the category of the last interacted item is selected. By comparing this category with those of other items, a Boolean vector is generated, which is then converted into binary form (with true replaced by 1 and false by 0). Finally, the diagonal mask matrix $\mathbf{M}$ is constructed, with the diagonal elements consisting of this binary vector. The matrix multiplication of the category mask matrix $\mathbf{M}$ with $\mathbf{u}_s$
yields a short-term interest representation $\mathbf{u}_c$, weighted by category information. This allows the model to focus more on the user's interest related to the target category within the current session.

\begin{equation}
\mathbf{u}_c = \mathbf{M}\,\mathbf{u}_s.
\end{equation}

Ultimately, the user's short-term interest is expressed as:
\begin{equation}
\mathbf{u}^S = \operatorname{concat}\!\left(\mathbf{u}_s, \mathbf{u}_c\right)
= [\mathbf{u}_s;\mathbf{u}_c].
\end{equation}
\subsection{Long-term Interest Encoder}
The encoder efficiently derives long-term interest representations for each user by integrating an attention-based encoder with a GRU model.

After the time-aware session segmentation layer, the user's historical behavior sequence is partitioned into $k$ sessions, with the first 
$k-1$ sessions utilized to extract long-term interests. For each historical session $S^{(n)}$, where $n \in \{1, \ldots, k-1\}$, the interest representation $\mathbf{h}_n$ is obtained through the attention encoder, as expressed by:
\begin{equation}
\mathbf{h}_n = \operatorname{Attention}\!\left(S^{(n)}\right),
\end{equation}
This process allows for the independent extraction of interest representations $\mathbf{h}_n$ for each historical session, resulting in a set of interest representations $\{\mathbf{h}_1, \dots, \mathbf{h}_{k-1}\}$ across $k-1$ sessions. 

\begin{figure}[h]
	\centering
	\includegraphics*[width=0.45\textwidth]{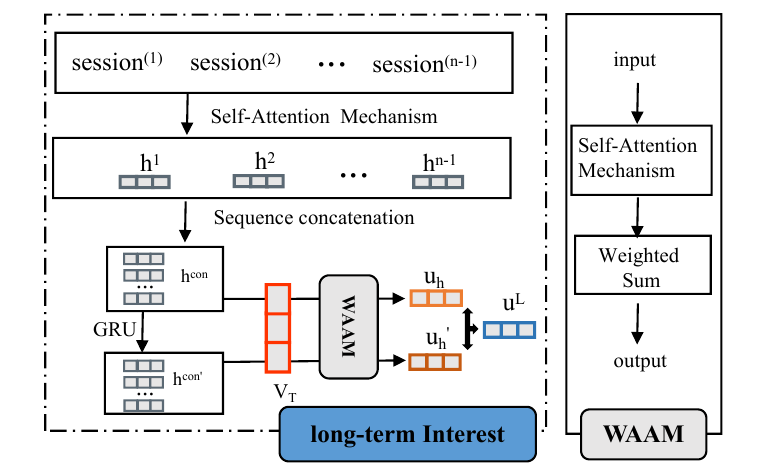}
	\caption{Long-term interest encoder}
	\label{fig:fig2}
\end{figure}

Next, we utilize Gated Recurrent Units (GRUs) \cite{dey2017gate} to model the long-term evolution of inter-session interests. While traditional RNNs are effective in time-series modeling, their low-parallelism architecture results in long-term dependency issues and inefficiencies when processing long sequences. Specifically, we encode the interest representation of each session using a GRU network, thereby obtaining a comprehensive representation of the user's long-term interests.
\begin{equation}
\big[\mathbf{h}_1', \mathbf{h}_2', \ldots, \mathbf{h}_{k-1}'\big]
=
\operatorname{GRU}\!\left(
\big[\mathbf{h}_1, \mathbf{h}_2, \ldots, \mathbf{h}_{k-1}\big]
\right),
\end{equation}

We employ an attention pooling mechanism to quantify the contribution of each session to the target item. Specifically, a soft alignment is established between each session representation $\mathbf{h}_i$ and the target item embedding $\mathbf{v}_T$. A learnable transformation matrix $\mathbf{W} \in \mathbb{R}^{d \times d}$ is applied to enable differentiated session contributions.
\begin{equation}
a_i =
\frac{
\exp\!\left(\mathbf{h}_i^{\top}\mathbf{W}\mathbf{v}_T\right)
}{
\sum_{j=1}^{k-1}
\exp\!\left(\mathbf{h}_j^{\top}\mathbf{W}\mathbf{v}_T\right)
},
\end{equation}
where $a_i$ denotes the attention weight of the $i$-th session, which are used to aggregate session representations into the long-term interest vector $\mathbf{u}_h$.
\begin{equation}
\mathbf{u}_h
=
\sum_{i=1}^{k-1} a_i \mathbf{h}_i .
\end{equation}

Similarly, we apply the same attention pooling operation to $\{\mathbf{h}_1',\dots, \mathbf{h}_{k-1}'\}$. Specifically, the weighted aggregation attention mechanism (WAAM) assigns
adaptive importance weights to inter-session representations, which are then
aggregated to obtain the long-term interest representation $\mathbf{u}_h'$.
Ultimately, the user's long-term interest is represented as:
\begin{equation}
\mathbf{u}^L=
\operatorname{concat}\!\left(\mathbf{u}_h,\mathbf{u}_h'\right)
=
[\mathbf{u}_h \, ; \, \mathbf{u}_h'] .
\end{equation}

\subsection{Adaptive Fusion and Model Prediction}
When predicting future user interactions, it is essential to jointly consider both long-term and short-term interests. Their relative importance varies over time: short-term interests tend to dominate when users repeatedly interact with similar items, whereas long-term preferences become more influential as users explore novel items. To capture this dynamic balance, we propose an adaptive aggregation mechanism that flexibly integrates long- and short-term interest representations according to the target item $\mathbf{v}_T$.
\begin{equation}
\alpha=\mathrm{\sigma}\!\left(
\boldsymbol{W}^q \mathrm{Concat}(\mathbf{u}^L,\mathbf{u}^S,\mathbf{v}_T)
+\boldsymbol{b}_q
\right),
\end{equation}

\begin{equation}
\mathbf{u}^{LS}=\alpha\mathbf{u}^L+(1-\alpha)\mathbf{u}^S,
\end{equation}

where $\sigma(\cdot)$ denotes the sigmoid function and $\mathbf{W}^q$ and $\mathbf{b}_q$ are learnable parameters. Here, $\alpha$ represents an adaptive fusion weight determined by the target item and the user's LS-term interests. Finally, we employ a two-layer MLP \cite{he2017neural} to predict the interaction score of the target item $\mathbf{v}_T$:

\begin{equation}
\hat{y}_{u,v}
=\sigma\!\left(
\operatorname{MLP}\!\left(
\mathbf{u}^{LS}, \mathbf{v}_T
\right)
\right).
\end{equation}

\subsection{Contrastive Learning}
Long-term interests model stable global preferences, whereas short-term interests capture recent behavioral tendencies. We supervise both encoders using averaged item embeddings from LS-term sessions to enhance dynamic interest modeling.

Specifically, the average embedding of items from the first $k-1$ sessions represents long-term interests, while that from the $k$-th session serves as the supervised short-term representation. Formally, given a user’s interaction sequence segmented into $k$ sessions ${\mathcal{S}^{(1)}, \ldots, \mathcal{S}^{(k)}}$, the supervised representations are defined as follows:
\begin{equation}
\begin{aligned}
\hat{U}^L
&= \frac{1}{t-(L-1)} \sum_{j=1}^{t-(L-1)} E\!\left(x_{j}^{u}\right),
\end{aligned}
\end{equation}

\begin{equation}
\begin{aligned}
\hat{U}^{S}
&= \frac{1}{L} \sum_{j=t-L+1}^{t} E\!\left(x_{j}^{u}\right),
\end{aligned}
\end{equation}
where $\mathcal{S}^{(n)} = \{x_{n,1}^u, \ldots, x_{n,L}^u\}$ denotes the set of items in the $n$-th session, $E(x)$ denotes the embedding representation of item $x$, $t$ represents the total number of historical interactions, and $L$ is the number of items in each session.

\begin{equation}\begin{gathered}
sim(u^L, \hat{U}^{L})>sim(u^L,\hat{U}^{S}), 
sim(\hat{U}^{L},u^L)>sim(\hat{U}^{L},u^S), \\
sim(u^S,\hat{U}^{S})>sim(u^S,\hat{U}^{L}), sim(\hat{U}^{S},u^S)>sim(\hat{U}^{S},u^L),
\end{gathered}\end{equation}
To achieve the above comparison learning objective, we use Triplet Loss for optimization, which is computed separately for LS-term interests. Formally, we capture the similarity of embeddings using Euclidean distance.
\begin{equation}
\label{eq:triplet}
f(a, p, n)
=
\max \left(
\| a - p \|_2^2
-
\| a - n \|_2^2
+
\alpha,
\; 0
\right),
\end{equation}
where $\|\cdot \|_2^2$ denotes the squared Euclidean distance, and $\alpha$ is a margin hyperparameter that enforces a sufficient separation between positive and negative pairs. The $\max(\cdot,0)$ operator ensures non-negativity of the loss by activating it only when the margin constraint is violated.
Specifically, we construct four triplet objectives by instantiating
$(a, p, n)$ with different combinations of LS-term
interest representations. Therefore, the loss function used to monitor interest is denoted as:

\begin{equation}
\begin{aligned}
\mathcal{L}_{\mathrm{con}}^{u}
&= f\!\left(u^L, \hat{U}^L, \hat{U}^S\right)
 + f\!\left(\hat{U}^L, u^L, u^S\right) \\
&\quad
 + f\!\left(u^S, \hat{U}^L, \hat{U}^S\right)
 + f\!\left(\hat{U}^S, u^S, u^L\right).
\end{aligned}
\end{equation}

Based on the settings of existing works \cite{yu2019adaptive},
we use the negative log-likelihood loss function as follows.
\begin{equation}
\mathcal{L}_{\mathrm{main}}^{u}
=
-\frac{1}{N}
\sum_{\upsilon \in O}
\left[
y_{u,\upsilon} \log \hat{y}_{u,\upsilon}
+
(1 - y_{u,\upsilon}) \log (1 - \hat{y}_{u,\upsilon})
\right],
\end{equation}

where $O$ denotes the training set consisting of one positive item and $n-1$ sampled negative items for each user.

Finally, our overall loss is defined as:
\begin{equation}
\mathcal{L}=\sum_{u=1}^{U}\left(\mathcal{L}_{\mathrm{main}}^{u}+\lambda\mathcal{L}_{\mathrm{con}}^{u}\right).\end{equation}

where $U$ denotes the total number of users.

\section{Experiments}
\begin{table}[!ht]
  \centering
  \caption{Statistics of the datasets.}
  \label{statistic}
  \footnotesize
  \setlength{\tabcolsep}{4pt}        
  \renewcommand{\arraystretch}{0.95}  
  \begin{tabular}{lrrrrr}
    \toprule
    Dataset & \multicolumn{1}{c}{\#Users} & \multicolumn{1}{c}{\#Items} & \multicolumn{1}{c}{\#Inte.} & \multicolumn{1}{c}{\#Cate.} & \multicolumn{1}{c}{\#Avg. Len} \\
    \midrule
    Taobao     & 42{,}966 & 103{,}204 & 2{,}089{,}165 & 2{,}936 & 48.62 \\
    Tmall      & 76{,}853 & 114{,}209 & 5{,}130{,}294 & 2{,}113 & 66.76 \\
    Cosmetics  & 2{,}515  & 5{,}288   & 109{,}580    & 309    & 43.57 \\
    \bottomrule
  \end{tabular}
\end{table}

\begin{table*}[t!]
  \centering
  \caption{Performance comparison.}
  \label{comparison}
  \footnotesize
  \setlength{\tabcolsep}{1pt}
  \renewcommand{\arraystretch}{0.9}
  \begin{tabular}{llccccccccc}
    \toprule
    Dataset & Model & AUC & GAUC & MRR & NDCG@2 & NDCG@5 & NDCG@10 & HIT@2 & HIT@5 & HIT@10 \\
    \midrule
    \multirow{12}{*}{Taobao}
      & NCF\cite{an2019neural}    & 0.7261 & 0.7356 & 0.2215 & 0.1689 & 0.2094 & 0.2423 & 0.1874 & 0.2786 & 0.3810 \\
      & DIN\cite{zhou2018deep}     & 0.8264 & 0.8442 & 0.4848 & 0.4429 & 0.4942 & 0.5212 & 0.4773 & 0.5910 & 0.6742 \\
      & CASER\cite{tang2018personalized}   & 0.8446 & 0.8491 & 0.4590 & 0.4150 & 0.4690 & 0.4976 & 0.4518 & 0.5714 & 0.6597 \\
      & GRU4REC\cite{hidasi2018recurrent} & 0.8580 & 0.8537 & 0.4639 & 0.4214 & 0.4732 & 0.4999 & 0.4573 & 0.5719 & 0.6545 \\
      & DIEN\cite{zhou2019deep}    & 0.8455 & 0.8441 & 0.4763 & 0.4364 & 0.4848 & 0.5092 & 0.4695 & 0.5767 & 0.6520 \\
      & SASREC\cite{kang2018self}  & 0.8419 & 0.8441 & 0.4897 & 0.4505 & 0.4995 & 0.5234 & 0.4834 & 0.5916 & 0.6555 \\
      & BERT4REC\cite{sun2019bert4rec}& 0.8503 & 0.8509 & 0.4773 & 0.4378 & 0.4880 & 0.5116 & 0.4737 & 0.5845 & 0.6576 \\
      \cmidrule(lr){2-11} 
      & SLIREC\cite{yu2019adaptive},  & 0.8410 & 0.8442 & 0.4386 & 0.3927 & 0.4465 & 0.4755 & 0.4271 & 0.5466 & 0.6362 \\
      & CLSR\cite{zheng2022disentangling}    & 0.8610 & 0.8585 & 0.4949 & 0.4560 & 0.5043 & 0.5291 & 0.4910 & 0.5977 & 0.6742 \\
      & LSIDN\cite{zhang2024denoising}   & 0.9039 & 0.9025 & 0.5194 & 0.4733 & 0.5359 & 0.5667 & 0.5167 & 0.6553 & 0.7500 \\
      \cmidrule(lr){2-11} 
    
      & SLSREC    & {\textbf{0.9076}} & \textbf{0.9066} & \textbf{0.5528} & \textbf{0.5109} & \textbf{0.5678} & \textbf{0.5957} & \textbf{0.5513} & \textbf{0.6769} & \textbf{0.7630} \\
      & Impro.   & \textbf{+0.41\%} & \textbf{+0.45\%} & \textbf{+6.43\%} & \textbf{+7.94\%} & \textbf{+5.95\%} & \textbf{+5.12\%} & \textbf{+6.69\%} & \textbf{+3.29\%} & \textbf{+1.73\%} \\
    \midrule
    \multirow{12}{*}{Tmall}
      & NCF\cite{an2019neural}    & 0.6655 & 0.7341 & 0.3632 & 0.3246 & 0.3529 & 0.3763 & 0.3359 & 0.3996 & 0.4725 \\
      & DIN\cite{zhou2018deep}     & 0.7988 & 0.8027 & 0.4218 & 0.3813 & 0.4200 & 0.4445 & 0.4020 & 0.4882 & 0.5639 \\
      & CASER\cite{tang2018personalized}   & 0.8138 & 0.8122 & 0.4037 & 0.3620 & 0.4028 & 0.4283 & 0.3851 & 0.4759 & 0.5551 \\
      & GRU4REC\cite{hidasi2018recurrent} & 0.7959 & 0.7983 & 0.4118 & 0.3704 & 0.4102 & 0.4351 & 0.3917 & 0.4803 & 0.5575 \\
      & DIEN\cite{zhou2019deep}    & 0.8404 & 0.8338 & 0.4279 & 0.3859 & 0.4248 & 0.4510 & 0.4063 & 0.4932 & 0.5742 \\
      & SASREC\cite{kang2018self}  & 0.8190 & 0.8134 & 0.4331 & 0.3914 & 0.4257 & 0.4529 & 0.4083 & 0.4853 & 0.5698 \\
      & BERT4REC\cite{sun2019bert4rec}& 0.8210 & 0.8164 & 0.4353 & 0.3981 & 0.4298 & 0.4516 & 0.4170 & 0.4875 & 0.5555 \\
      \cmidrule(lr){2-11} 
      & SLIREC\cite{yu2019adaptive}  & 0.8003 & 0.8014 & 0.4099 & 0.3682 & 0.4080 & 0.4330 & 0.3893 & 0.4780 & 0.5556 \\
      & CLSR\cite{zheng2022disentangling}    & 0.8191 & 0.8144 & 0.4261 & 0.3857 & 0.4260 & 0.4500 & 0.4072 & 0.4970 & 0.5714 \\
      & LSIDN\cite{zhang2024denoising}   & 0.8534 & 0.8589 & 0.4624 & 0.4174 & 0.4646 & 0.4930 & 0.4441 & 0.5494 & 0.6373 \\
      \cmidrule(lr){2-11} 
      & SLSREC    & \textbf{0.8599} & \textbf{0.8603} & \textbf{0.4698} & \textbf{0.4252} & \textbf{0.4711} & \textbf{0.5000} & \textbf{0.4509} & \textbf{0.5531} & \textbf{0.6425} \\
      & Impro.   & \textbf{+0.76\%} & \textbf{+0.16\%} & \textbf{+1.60\%} & \textbf{+1.87\%} & \textbf{+1.40\%} & \textbf{+1.42\%} & \textbf{+1.53\%} & \textbf{+0.67\%} & \textbf{+0.82\%} \\
    \midrule
    \multirow{12}{*}{Cosmetics}
      & NCF\cite{an2019neural}     & 0.6333 & 0.6487 & 0.1708 & 0.1251 & 0.1578 & 0.1835 & 0.1397 & 0.2123 & 0.2915 \\
      & DIN\cite{zhou2018deep}     & 0.8520 & 0.8634 & 0.4881 & 0.4468 & 0.5029 & 0.5314 & 0.4906 & 0.6172 & 0.7052 \\
      & CASER\cite{tang2018personalized}   & 0.8404 & 0.8528 & 0.4697 & 0.4210 & 0.4835 & 0.5147 & 0.4620 & 0.5996 & 0.6964 \\
      & GRU4REC\cite{hidasi2018recurrent} & 0.8691 & 0.8725 & 0.4629 & 0.4150 & 0.4782 & 0.5074 & 0.4576 & 0.5974 & 0.6876 \\
      & DIEN\cite{zhou2019deep}    & 0.8555 & 0.8559 & 0.4797 & 0.4366 & 0.4882 & 0.5183 & 0.4719 & 0.5853 & 0.6799 \\
      & SASREC\cite{kang2018self}  & 0.8667 & 0.8695 & 0.5018 & \textbf{0.4703} & \textbf{0.5234} & 0.5492 & 0.5038 & 0.6221 & 0.6999 \\
      & BERT4REC\cite{sun2019bert4rec}& 0.8468 & 0.8533 & 0.4690 & 0.4200 & 0.4809 & 0.5115 & 0.4510 & 0.5856 & 0.6792 \\
      \cmidrule(lr){2-11} 
      & SLIREC\cite{yu2019adaptive}  & 0.8303 & 0.8374 & 0.4220 & 0.3700 & 0.4260 & 0.4596 & 0.4004 & 0.5226 & 0.6260 \\
      & CLSR\cite{zheng2022disentangling}    & 0.8723 & 0.8762 & 0.4802 & 0.4355 & 0.4919 & 0.5245 & 0.4785 & 0.6018 & 0.7030 \\
      & LSIDN\cite{zhang2024denoising}   & 0.8620 & 0.8762 & 0.4738 & 0.4206 & 0.4911 & 0.5237 & 0.4620 & 0.6183 & 0.7195 \\
      \cmidrule(lr){2-11} 
      & SLSREC    & \textbf{0.8734} & \textbf{0.8793} & \textbf{0.5066} & 0.4648 & 0.5212 & \textbf{0.5514} & \textbf{0.5083} & \textbf{0.6348} & \textbf{0.7283} \\
      & Impro.   & \textbf{+0.13\%} & \textbf{+0.35\%} & \textbf{+0.96\%} & -1.17\% & -0.42\% & \textbf{+0.4\%} & \textbf{+0.8\%} & \textbf{+2.04\%} & \textbf{+1.22\%} \\
    \bottomrule
  \end{tabular}
\end{table*}

In this section, we conduct experiments to demonstrate the effectiveness of the proposed model, SLSRec.
Specifically, we aim to answer the following research questions.
 \begin{itemize}
    \item RQ1: How does our framework perform in terms of overall performance compared to state-of-the-art recommendation models, and does interest modeling at different time scales more accurately capture dynamic changes in user interests?
    \item RQ2: How do the different components influence the performance of SLSRec?
    \item RQ3: How do hyperparameters influence the effectiveness of SLSRec?
 \end{itemize}

\textbf{Datasets.} We conducted experiments on publicly available e-commerce datasets from Taobao\footnote{\url{https://tianchi.aliyun.com/dataset/649}}, 
Tmall\footnote{\url{https://tianchi.aliyun.com/dataset/140281}}, 
Cosmetics\footnote{\url{https://www.kaggle.com/datasets/mkechinov/ecommerce-events-history-in-cosmetics-shop?select=2020-Jan.csv}}, encompassing various user behaviors such as clicks, collections, add-to-cart actions, and purchases recorded at different timestamps.Detailed statistics of the datasets are summarized in Table~\ref{statistic}.
We conduct experiments on three real-world e-commerce datasets: Taobao, Tmall, and Cosmetics. For the Taobao dataset, user--item interactions were collected from November~25 to December~4, 2017. Interactions occurring before December~1 were used for training, while those from December~2 to December~4 were used for validation and testing to ensure a strict temporal evaluation protocol.
Detailed preprocessing procedures for the Tmall and Cosmetics datasets are provided in the appendix C.1.



\textbf{Baselines and Metrics.} 
To evaluate the effectiveness of SLSRec, we compare it with representative baselines covering different interest modeling paradigms, including holistic models (NCF \cite{an2019neural}, DIN \cite{zhou2018deep}, Caser \cite{tang2018personalized}, GRU4Rec \cite{hidasi2018recurrent}, and DIEN \cite{zhou2019deep}), Transformer-based sequential models (SASRec \cite{kang2018self} and BERT4Rec \cite{sun2019bert4rec}), and long-short term interest models (SLi-Rec \cite{yu2019adaptive}, CLSR \cite{zheng2022disentangling}, and LSDIN \cite{zhang2024denoising}).

We evaluate all methods using standard accuracy metrics (AUC, GAUC \cite{zhou2018deep}, Hit@K) and ranking metrics (MRR and NDCG@K \cite{jarvelin2002cumulated}). For fair comparison, all models are implemented using a unified TensorFlow-based framework \cite{argyriou2020microsoft}, with identical initialization strategies and early stopping criteria following \cite{zhang2024denoising}. Unless otherwise specified, the batch size, learning rate, and maximum sequence length are set to 500, 0.001, and 50, respectively. The hyperparameter $\lambda$ is set to 0.2 for Taobao and Tmall, and 0.1 for Cosmetics, while the time threshold $\omega$ is set to 90 minutes, one day, and 30 minutes for Taobao, Tmall, and Cosmetics, respectively.

\subsection{Overall Performance Comparison (RQ1)}

In Table ~\ref{comparison}, we detail the performance of all models on different metrics and summarize several key findings accordingly.


\textbf{Incorporating multiple time scales in interest modeling consistently outperforms holistic approaches.}User interests exhibit inherently multi-scale dynamics, which cannot be adequately captured by a single unified representation. As a result, holistic interest modeling often fails to simultaneously reflect short-term responsiveness and long-term stability, leading to suboptimal performance. Experimental results on three real-world datasets confirm that multi-scale models consistently surpass the strongest holistic baseline across all metrics, with AUC improvements of up to 5.78\%, highlighting the effectiveness of multi-scale interest modeling.

\textbf{Proper fusion of LS-term user interests is essential for enhancing recommendation performance.} SLi-Rec fails to yield performance improvements and instead introduces additional model complexity. Experimental results on all three datasets demonstrate that most holistic modeling approaches outperform SLi-Rec, indicating that improper fusion of long- and short-term interests can adversely impact user behavior modeling. In contrast, SLSRec employs an attention mechanism to adaptively integrate LS-term interests, leading to significant improvements in recommendation accuracy.

\textbf{Modeling user interests at multiple time scales enables more accurate capture of dynamic preference evolution by balancing short-term responsiveness and long-term stability.}
On the Taobao and Tmall datasets, multi-scale models (e.g., LSDIN and SLSRec) consistently outperform the holistic baseline (CLSR), achieving MRR improvements of 4.95\% and 11.69\% on Taobao, and 8.51\% and 10.26\% on Tmall, respectively. Moreover, these models yield over 10\% gains in HIT@10, demonstrating their superiority in ranking performance. These results indicate that explicitly distinguishing short-term and long-term behaviors is critical for effective recommendation. By hierarchically modeling user interests across time scales, multi-scale approaches better capture interest dynamics while mitigating noise from long-term behaviors, thereby improving both accuracy and robustness in real-world recommendation scenarios.

Finally, the experimental results show that the proposed model SLSRec outperforms all baseline models in all experimental settings. 

\subsection{Ablation Study (RQ2)}

\begin{table}[t!]
  \centering
  \caption{Ablation study results on three datasets.}
  \label{ablation}
  \footnotesize
  \setlength{\tabcolsep}{5pt}
  \renewcommand{\arraystretch}{0.9}
  \begin{tabular}{llcccc}
    \toprule
    Dataset & Model & AUC & MRR & NDCG@2 & HIT@2 \\
    \midrule
    \multirow{5}{*}{Taobao}
      & SLSREC      & \textbf{0.9076} & \textbf{0.5528} & \textbf{0.5109} & \textbf{0.5513} \\
      & w/o CL     & 0.9026 & 0.5316 & 0.4874 & 0.5279 \\
      & w/o cate   & 0.8985 & 0.5167 & 0.4710 & 0.5119 \\
      & w/o long   & 0.8639 & 0.4992 & 0.4588 & 0.4936 \\
      & w/o short  & 0.6559 & 0.1624 & 0.1064 & 0.1238 \\
    \midrule
    \multirow{5}{*}{Tmall}
      & SLSREC & \textbf{0.8599} & \textbf{0.4698} & \textbf{0.4252} & \textbf{0.4509} \\
      & w/o CL     & 0.8514 & 0.4680 & 0.4244 & 0.4502 \\
      & w/o cate   & 0.8559 & 0.4676 & 0.4233 & 0.4483 \\
      & w/o long   & 0.8382 & 0.4497 & 0.4050 & 0.4303 \\
      & w/o short  & 0.5419 & 0.1209 & 0.0857 & 0.0943 \\
    \midrule
    \multirow{5}{*}{Cosmetics}
      & SLSREC  & \textbf{0.8734} & \textbf{0.5066} & \textbf{0.4648} & \textbf{0.5083} \\
      & w/o CL     & 0.8661 & 0.4939 & 0.4494 & 0.4851 \\
      & w/o cate   & 0.8683 & 0.4956 & 0.4467 & 0.4873 \\
      & w/o long   & 0.8427 & 0.4747 & 0.4327 & 0.4785 \\
      & w/o short  & 0.5249 & 0.1056 & 0.0710 & 0.0836 \\
    \bottomrule
  \end{tabular}
\end{table}


We perform an ablation study by selectively removing the contrastive learning module, category-aware masking, long-term encoder, or short-term encoder from SLSRec.

As shown in Table~\ref{ablation}, removing any key component of SLSRec results in notable performance degradation, demonstrating the necessity of each module. Among them, the short-term interest encoder contributes most significantly to overall performance. Removing this module causes drastic drops in AUC on Taobao, Tmall, and Cosmetics (27.73\%, 36.98\%, and 39.90\%, respectively), along with severe declines in NDCG@2 (70.62\%, 74.27\%, and 79.16\%). These results highlight the crucial role of short-term interest modeling in capturing users’ immediate preferences.

Eliminating the long-term interest encoder also leads to consistent performance degradation, with AUC decreasing by 4.81\%, 2.52\%, and 3.52\% on the three datasets, and NDCG@2 dropping by 10.20\%, 4.27\%, and 6.30\%. This confirms that long-term interest modeling provides complementary benefits and that jointly modeling LS-term interests is essential for capturing interest dynamics.

In addition, removing the category-masking module negatively affects performance across multiple metrics. For example, NDCG@2 decreases by 7.81\% and 3.89\% on Taobao and Cosmetics, while HIT@2 drops by 7.15\% and 4.13\%, respectively. This indicates that category-aware short-term modeling helps the model focus on session-relevant interests.

Finally, while the removal of the contrastive learning component results in a relatively minor decline in model performance, its impact remains observable. This suggests that even when the model exhibits strong overall performance, contrastive learning continues to play a beneficial role by facilitating the optimization of LS-term interest representations. 

\subsection{Hyper-parameter Study (RQ3)}
\begin{figure}[htbp]  
    \centering
    \begin{subfigure}{0.49\linewidth}
        \centering
        \includegraphics[width=\linewidth]{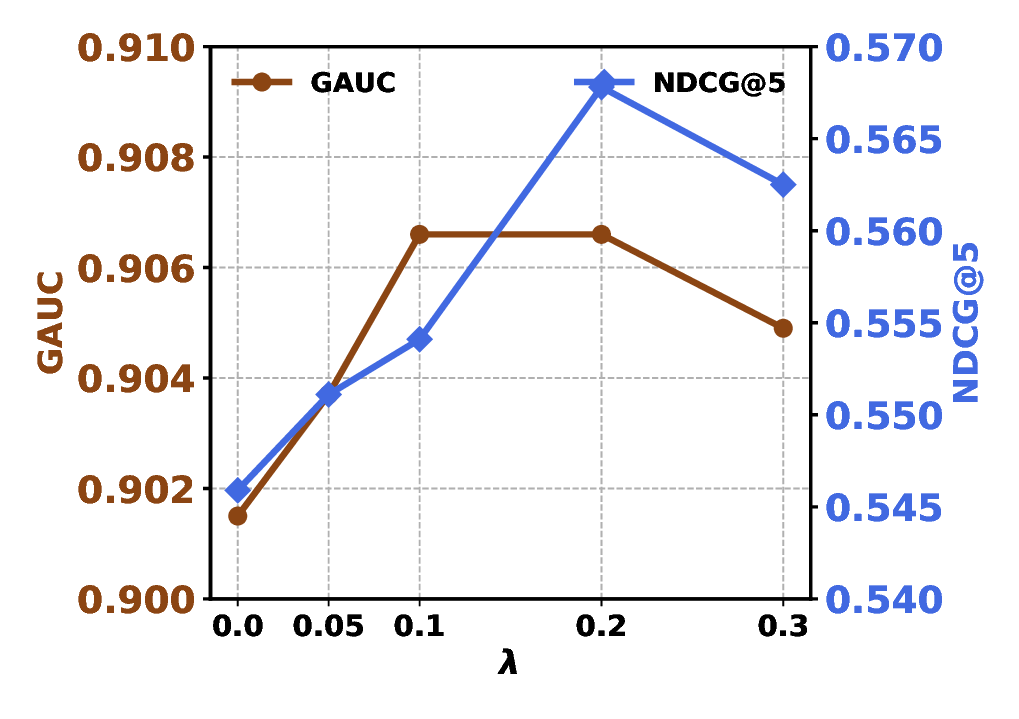}
        \caption{Taobao}
        \label{fig:sub_a1}
    \end{subfigure}
    \hfill
    \begin{subfigure}{0.49\linewidth}
        \centering
        \includegraphics[width=\linewidth]{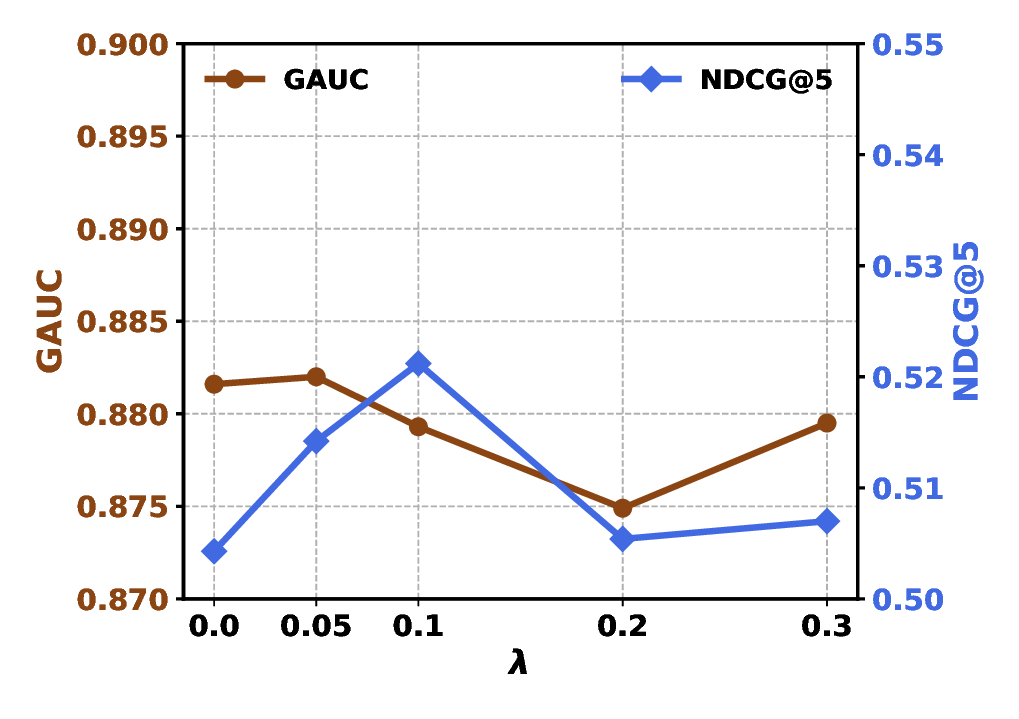}
        \caption{Cosmetics}
        \label{fig:sub_b1}
    \end{subfigure}

    \caption{$\lambda$ analyses of SLSRec on the Taobao and Cosmetics dataset}
    \label{fig:main1}
\end{figure}

We investigate the impact of two key hyper-parameters on model performance on Taobao and Cosmetics datasets:
Contrast loss weight $\lambda$ controls the strength of contrast regularization. If $\lambda$ is too large, it may conflict with the main learning objective; if it is too small, it may lead to an insufficient regularization effect, thus failing to improve the model performance effectively. As shown in Figure \ref{fig:sub_a1}, the experimental results show that the overall performance of the model is best when $\lambda$ = $0.2$ for the Taobao dataset. For the Comestic dataset, the best performance is when $\lambda$ is equal to $0.1$.

\begin{figure}[htbp]  
    \centering
    \begin{subfigure}{0.49\linewidth}
        \centering
        \includegraphics[width=\linewidth]{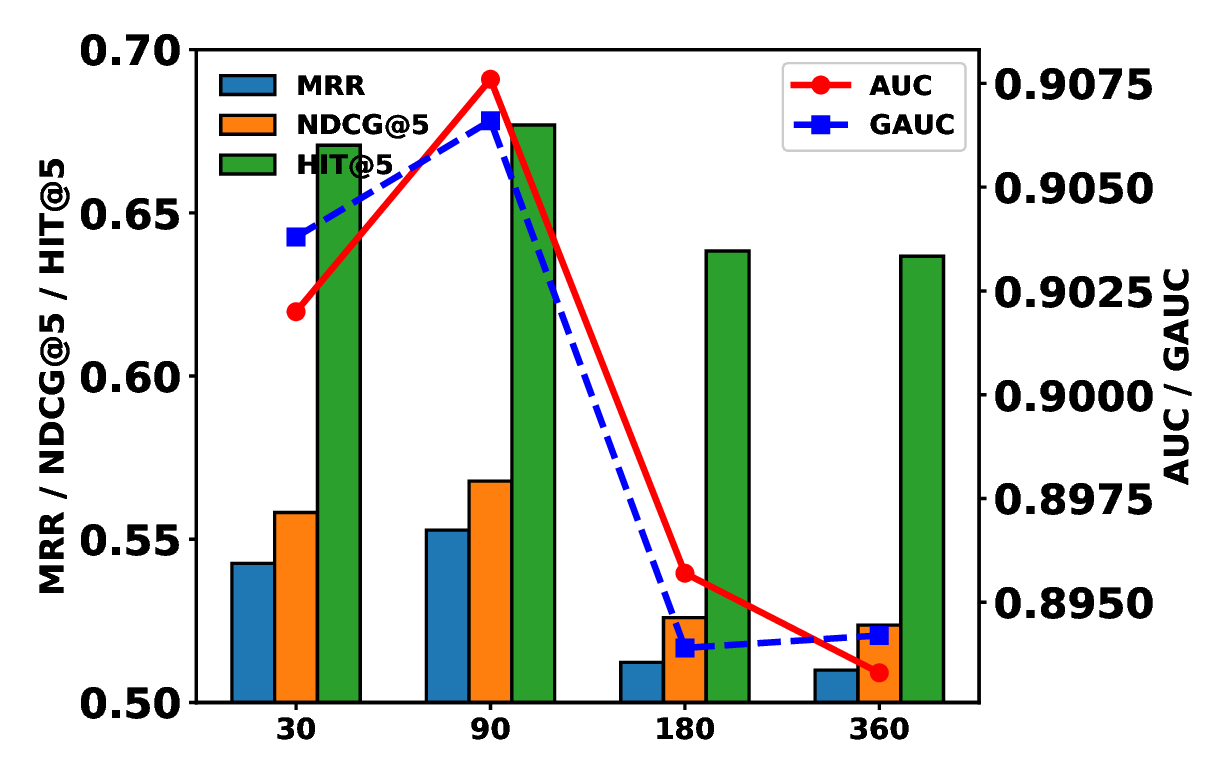}
        \caption{Taobao}
        \label{fig:sub_a}
    \end{subfigure}
    \hfill
    \begin{subfigure}{0.49\linewidth}
        \centering
        \includegraphics[width=\linewidth]{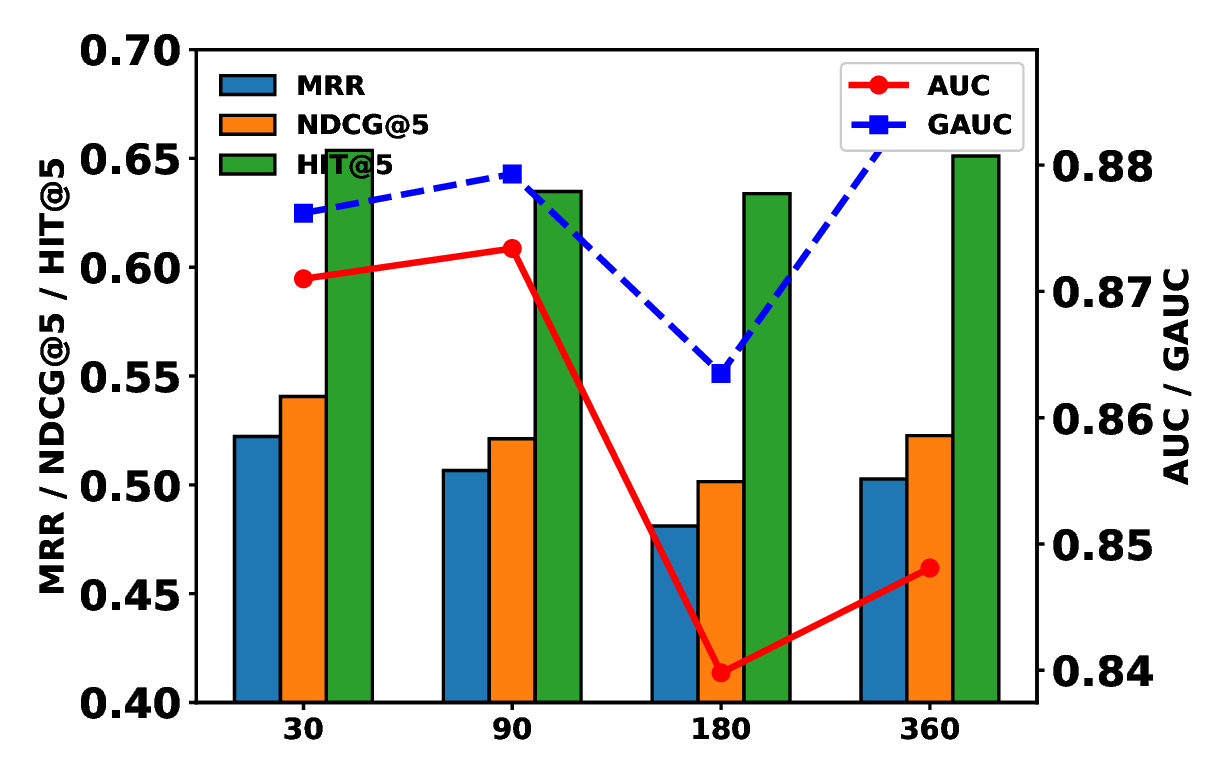}
        \caption{Cosmetics}
        \label{fig:sub_b}
    \end{subfigure}

    \caption{Impact of $\omega$ variation on the effectiveness of SLSRec across Taobao and Cosmetics datasets}
    \label{fig:main}
\end{figure}

Session Segmentation Threshold $\omega$ controls how user behavior sequences are segmented into sessions. A small $\omega$ may cause over-segmentation, resulting in overly short sessions that fail to capture sufficient user intent, while a large $\omega$ may include irrelevant behaviors within a session and introduce noise. As shown in Fig.~\ref{fig:sub_a}, the optimal $\omega$ is 90 minutes for Taobao and 30 minutes for Cosmetics; deviations from these values consistently lead to performance degradation.
\section{Conclusion}
In this paper, we propose \textbf{SLSRec}, a contrastive learning-based framework for sequential recommendation that explicitly models users’ LS-term interests through segmented interaction sequences. To enhance the expressiveness of interest representations, SLSRec introduces a self-supervised contrastive mechanism to disentangle long-term preferences and short-term intentions, enabling more accurate modeling of users’ evolving behaviors. Extensive experiments conducted on multiple benchmark datasets demonstrate that SLSRec consistently outperforms state-of-the-art recommendation methods in terms of both effectiveness and robustness. These results underscore the effectiveness of contrastive learning for interest disentanglement and highlight the importance of explicitly modeling multi-scale user interests in sequential recommendation.

\bibliographystyle{named}
\bibliography{ijcai26}

\end{document}